\documentclass[12pt]{article}

\usepackage[dvips]{graphicx}

\newcommand{\lsim}{\raisebox{-4pt}{$\,\stackrel{\textstyle
                                                         <}{\sim}\,$}}
\newcommand{\gsim}{\raisebox{-4pt}{$\,\stackrel{\textstyle
                                                         >}{\sim}\,$}}
\newcommand{\nn}{\nonumber}
\newcommand{\be}{\begin{equation}}
\newcommand{\ee}{\end{equation}}
\newcommand{\ba}{\begin{eqnarray}}
\newcommand{\ea}{\end{eqnarray}}
\newcommand{\req}[1]{(\ref{#1})}
\def\={\,=\,}
\newcommand{\ci}[1]{\cite{#1}}

\def\mev{~{\rm MeV}}
\def\gev{~{\rm GeV}}

\def\ale{\alpha_{\rm em}}
\def\als{\alpha_{\rm s}}
\def\eps{\epsilon}

\def\LQCD{\Lambda_{\rm QCD}}

\def\muF{\mu^2_F}

\newcommand{\tw}{\textwidth}
\def\vk{{\bf k}_{\perp}}

\def\vb0{{\bf b}_0}

\newcommand{\sla}{\hspace*{-0.030\tw}/}
\newcommand{\Sla}{\hspace*{-0.020\tw}/}
\newcommand{\da}{{distribution amplitude}}
\newcommand{\wf}{wave function}

\newcommand{\ov}[1]{\overline#1}

\def\={\,=\,}

\begin{document} 
\thispagestyle{empty}
\begin{flushright}
WUB/16-08\\
January, 05  2017\\[20mm]
\end{flushright}

\begin{center}
{\Large\bf A study of the $\gamma^*-f_{0}(980)$ transition form factors}
\vskip 10mm

P.\ Kroll \footnote{Email:  pkroll@uni-wuppertal.de}
\\[1em]
{\small {\it Fachbereich Physik, Universit\"at Wuppertal, D-42097 Wuppertal,
Germany}}\\
and
{\small {\it Institut f\"ur Theoretische Physik, Universit\"at
    Regensburg, \\D-93040 Regensburg, Germany}}\\

\centerline{revised version}
\end{center}
\vskip 5mm 
\begin{abstract}
The $\gamma^*-f_{0}(980)$ transition form factors are calculated within the QCD factorization
framework. The $f_0$-meson is assumed to be mainly generated through its $s\bar{s}$ Fock
component. The corresponding spin wave function of the $f_{0}(980)$ meson is constructed 
and, combined with a model light-cone wave function for this Fock component, used in the
calculation of the form factors. In the real-photon limit the results for the transverse 
form factor are compared to the large-momentum-transfer data measured by the BELLE 
collaboration recently. It turns out that, for the momentum-transfer range explored by 
BELLE, the collinear approximation does not suffice, power corrections to it, modeled as 
quark transverse moment effects, seem to be needed. Mixing of the $f_0$ with the 
$\sigma(500)$ is also briefly discussed.
\end{abstract} 
\section{Introduction}
\label{sec:intro}
Recently the BELLE collaboration \ci{belle15} has measured the 
cross section for $\gamma^*\gamma \to \pi^0\pi^0$ for large photon 
virtuality, $Q_1^2$, and small energy in the $\gamma^*\gamma$ center-of-mass
system. From these data the photon-meson transition form factors have been 
extracted for the scalar, $f_0(980)$, and tensor, $f_2(1270)$, mesons for 
$Q_1^2 \lsim 30\,\gev^2$. These transition form factors are similar to those
for the pseudoscalar mesons which have been extensively studied by both
experimentalists and theoreticians. In Ref.\ \ci{schuler97} the $\gamma -f_0$
and the $\gamma -f_2$ form factors have been investigated within the NRQCD
factorization framework \ci{bodwin95}, in which relativistic corrections and 
higher Fock state contributions are suppressed by powers of the relativistic 
velocity of the quarks in the meson, i.e.\ up to some minor modifications, the 
light mesons are treated like heavy Quarkonia. Super-convergence relations have 
been derived in \ci{pascalutsa12}  and shown to provide constraints on the 
$\gamma - f_2$ transition form factor. The latter form factor has also been
studied within the framework of collinear factorization \ci{braun16}. A 
phenomenological model for this form factor is discussed in \ci{achasov15}. 
The process $\gamma^*\gamma \to \pi\pi$ has been discussed in the framework of
generalized distribution amplitudes, time-like versions of generalized parton 
distributions \ci{diehl-pire}. In this paper the interest is focused on the 
$\gamma  - f_0$ transition form factor. 

The $f_0(980)$ meson is a complicated system whose nature is not yet fully understood. 
Its peculiar properties have led to many speculations about its quark content. 
A comparison of the partial widths for the $f_0$  decays into pairs of pions and 
Kaons \ci{PDG} under regard of the respective phase spaces reveals that the matrix 
element for $f_0\to K^+K^-$ is much larger than that for $f_0\to \pi^+\pi^-$. Thus,
if the $f_0$ is viewed as a quark-antiquark state, it is dominantly an $s\bar{s}$ state. 
The comparison of the branching ratios for the radiative decays of the $\phi$-meson
into the $f_0$ and $\pi^0$ leads to the same conclusion. However, the $f_0$-meson is not 
a pure $s\bar{s}$ state as is, for instance, obvious from the decay widths for 
$J/\Psi\to f_0\omega$ and $J/\Psi\to f_0\phi$. This fact is interpreted as $f_0 - \sigma(500)$
mixing. Detailed phenomenological analyses of $f_0-\sigma$ mixing in various decay processes 
\ci{ochs,cheng05,stone13,zhang16} revealed two ranges for the mixing angle, $\varphi$,
\be
 (25 - 40)^\circ \hspace*{0.1\tw} (140 - 165)^\circ
\label{eq:mixing}
\ee
A light scalar glueball may affect this result \ci{ochs}.   

As an alternative to the quark-antiquark interpretation other authors \ci{jaffe03,maiani04}
have  suggested a tetraquark configuration for the $f_0$-meson. This appears as a natural 
explanation for the fact that the $a_0(980)$ and the $f_0$ mesons are degenerate in mass and 
are the heaviest particles of the lightest scalar-meson nonet. For the tetraquark interpretation
there seems to be no $f_0-\sigma$ mixing \ci{stone13}. The drawback of this picture is that
the two-pion decay of the $f_0$ is too small as compared to experiment whereas
the $a_0\to\eta\pi$ is too large. In \ci{hooft} it has been suggested that the lightest 
scalar-meson nonet, considered as tetraquarks states, mixes with the scalar-meson nonet with 
masses around 1200 MeV under the effect of the instanton force. The latter nonet is believed 
to have a predominant $q\bar{q}$ structure. This mixing leads to a better description of the 
light scalar-meson decays. The $f_0$ may also have a substantial $K\bar{K}$ molecule
component \ci{baru03}. It goes without saying that the real $f_0$-meson is a superposition
of all these configurations.

The goal of the present paper is the calculation of the $\gamma^* - f_0$ transition form 
factors at large photon virtualities. For this calculation the pQCD framework  developed 
by Brodsky and Lepage \ci{brodsky-lepage79} is utilized in which the process is factorized 
in a perturbatively calculable hard subprocess (here $\gamma^*\gamma^*\to q\bar{q}$) 
and a soft hadronic matrix element, parametrized as a light-cone \wf, which is under control
of soft, long-distance QCD. As any hadron the $f_0$-meson possesses a Fock decomposition 
\ci{BHL} starting with the simple quark-antiquark components  
\ba
|f_0;p\rangle &=& \sum_\beta \int [d\tau]_2[d^2{\bf k}_\perp]_2 
                 \Psi_{2,\beta}(\tau,{\bf k}_\perp) |q\bar{q},\beta;k_1,k_2\rangle \nn\\
         &&  +\; \textrm{higher Fock states}
\label{eq:Fock}
\ea
where $\Psi_{2,\beta}$ is the light-cone \wf{} of the $q\bar{q}$ Fock state; the index $\beta$
labels its decomposition in flavor, color and helicity. The integration measures are defined by
\ba
[d\tau]_2   &=& d\tau_1d\tau_2\,\delta(1-\tau_1-\tau_2)\,, \nn\\  
{[d^2 {\bf k}_{\perp}]}_2&=& \frac{d^2{\bf k}_{\perp 1}d^2{\bf k}_{\perp 2}}{16\pi^3}\,
              \delta^{(2)}({\bf k}_{\perp 1}+{\bf k}_{\perp 2}-{\bf p}_\perp)\,. 
\ea
In the photon-photon interactions at large photon virtualities the $f_0$-meson is generated through 
its lowest Fock components, mainly the $s\bar{s}$ one. As can be shown \ci{brodsky-lepage79} the 
hard generation of the $f_0$ through higher Fock components is suppressed by inverse powers of the 
photon virtuality and is therefore neglected.  Once the meson is produced it gets dressed by 
fluctuations into higher Fock components under the effect of long-distance QCD. The calculation of 
the $\gamma^*-f_0$ transition form factors is similar to the one of the photon-pseudoscalar-meson 
form factors \ci{brodsky-lepage79}. The latter calculation is to be generalized in such a way that 
also hadrons with non-zero orbital angular between their constituents can be treated.

The paper is organized as follows: In the next section the spin part of the light-cone \wf{},
termed the spin wave function, of the $f_0$ is constructed assuming that this mesons is an 
$s\bar{s}$ state. In Sect.\ 2.1 the collinear reduction of the spin \wf{} is discussed and, in 
Sect.\ 2.2, an example of a light-cone \wf{} of the $f_0$ is introduced and compared to the 
twist-2 and 3 \da s. The $\gamma^*-f_0$ transition form factors are defined in Sect.\ 3.1, 
followed by a LO perturbative calculation within the modified perturbative approach in 
which quark transverse degrees of freedom are retained (Sect 3.2). Numerical results for 
the form factors in the real-photon limit are given in Sect. 4.1 and compared to the 
BELLE data. Some comments on the behavior of the $\gamma^* - f_0$ form factors are 
presented in Sect.\ 4.2. Finally, the summary will be given in Sect.\ 5.

\section{The spin \wf{} of the $f_0$-meson}
\label{sec:spin-wf}
For the description of the hadron the light-cone approach is used which enables one
to completely separate the dynamical and kinematical features of the Poincar\'e invariance
\ci{dirac,leutwyler}. The overall motion of the hadron is decoupled from the internal
motion of the constituents, i.e.\ the light-cone \wf{} of the hadron, $\Psi$, is independent
of the hadron's momentum and is invariant under the kinematical Poincar\'e transformations
( boosts along and rotations around the 3-directions as well as transverse boosts).
Hence, $\Psi$ is determined if it is known at rest. The $s\bar{s}$ Fock component given in 
\req{eq:Fock}, is split in a spin part (hereafter denoted as spin \wf{}) and a reduced
light-cone \wf{}, $\Psi_0$, which represents the full, soft \wf{}, $\Psi$, with a factor 
$K^\mu$ removed from it. As discussed in detail in Ref.\ \ci{bolz94}
the covariant spin \wf{} can be constructed starting from the observation \ci{dziembowski}
that, in zero binding energy approximation, an equal-time hadron state (in the spin basis)
in the constituent center-of-mass frame equals the (helicity) light-cone state at rest.
Consequently, one can use the standard $ls$ coupling scheme in order to couple quark and
antiquark to a state of given spin and parity. On boosting the results to a frame 
with arbitrary hadron momentum one easily reads off the covariant spin \wf{}~\footnote{
 In \ci{bolz94} this method has been applied for instance in a calculation of the
 $\pi - a_1(1260)$ form factors.}.  

Since the $f_0(980)$-meson is  a $J^{PC}= 0^{++}$ state the quark and antiquark have to 
couple in a spin-1 state and one unit of orbital angular momenta is required~\footnote{
In spectroscopy notation the valence Fock component of $f_0$-meson is a $^3P_0$ state.}.
The $ls$ coupling scheme leads to the following ansatz for the spin \wf{} of a final state
meson in its rest frame \ci{bolz94,hussain-koerner} ($\bar{S}_0=\gamma_0S^\dagger\gamma_0$) 
\be
\bar{S}_0\=\sum_{m,\mu_1,\mu_2}k\sqrt{4\pi}Y^*_{1m}({\bf k}/k)
      \left({1/2\atop \mu_1}{1/2 \atop \mu_2}\left|{1 \atop \mu_s}\right)\right.
\left({1 \atop \mu_s}{1 \atop m}\left|{0_{\phantom{s}} \atop 0_{\phantom{s}} }\right) \right.
                             v(\hat{p}_2,\mu_2)\bar{u}(\hat{p}_1,\mu_1)\,.
\label{eq:ls-scheme}
\ee
Note that $\mu_1, \mu_2$ denote spin components and $v, \bar{u}$ are equal-t
spinors here. In the meson's rest frame the meson and the constituent 
momenta read
\be
\hat{p}^\mu\=(M_0,{\bf 0})\,, \qquad \hat{p}_1^\mu\=(m_1,{\bf k})\,, \qquad 
       \hat{p}_2^\mu\=(m_2,-{\bf k})\,.
\ee
where ${\bf k}$ is the three-momentum part of the relative momentum of quark and antiquark
\be 
     {\bf k} \=\frac12\big(\hat{{\bf p}}_1 - \hat{{\bf p}}_2\big)\,.
\label{eq:K}
\ee
In order to retain a covariant formulation, the four-vector 
$K=(0,{\bf k})$ is introduced~\footnote{
  As discussed in \ci{bolz94} each unit of orbital angular momentum will be represented by
  $$K_\perp^\mu = K^\mu - \hat{v}\cdot K \,\hat{v}^\mu $$
  where $\hat{v}^\mu=\hat{p}^\mu/M_0=(1,{\bf 0})$ is the  velocity 4-vector.
  In the rest frame clearly $K_\perp \to (0,{\bf k})$ and one has the appropriate 
  object transforming as a 3-vector under $O(3)$. Thus, $K^\mu$ introduced in the line
  after \req{eq:K}, is strictly speaking $K_\perp^\mu$.}. 
As is customary in the parton model, the binding energy is neglected and the constituents 
are considered as quasi on-shell particles. That possibly crude approximation can be achieved
by putting the minus components of the constituents to zero. Hence, $k_3=0$ and our
relative vector reduces to 
\be
  K\=\big[0\, 0\, \vk \big]\,.
\label{eq:K4}
\ee
In this case the spin wave function \req{eq:ls-scheme} reads
\be
\bar{S}_0\= \frac1{\sqrt{2}} \Big[ k_{\perp +}v(\hat{p}_2,+)\bar{u}(\hat{p}_1,+)
                                  -k_{\perp -}v(\hat{p}_2,-)\bar{u}(\hat{p}_1,-) \Big]
\ee
where $k_{\perp \pm}=k_{\perp 1} \pm i k_{\perp 2}$.
This spin wave function is of the same type as is discussed in \ci{ji-ma03} for the $l=1$ 
Fock components of $\rho$ and $\pi$-mesons.

In the infinite momentum frame (IMF), obtained by boosting the meson rest frame momenta along 
the 3-direction, $p\cdot K=0$ holds and the quark and antiquark momenta are parametrized 
as~\footnote{
    In \ci{neubert} the parton momenta are parametrized as 
    $$ p_1=\tau p + K + \frac{k^2_\perp}{2\tau p\cdot \bar{p}} \bar{p}\,,  
\qquad   p_2=\bar{\tau} p - K + \frac{k^2_\perp}{2\bar{\tau}p\cdot \bar{p}} \bar{p} $$
    where $\bar{p}$ is a light-like vector whose 3-component points in the opposite
    direction of ${\bf p}$. For this parametrization momentum conservation
    only holds up to corrections of order $k^2_\perp/p$. It however also leads to the
    spin \wf{} \req{eq:spin-wf} up to corrections of order $k^3_\perp/M_0$.}
\be
p_1\=\tau p + K\,,  \qquad   p_2\=\bar{\tau} p - K
\label{eq:parton-momenta}
\ee
where $\bar{\tau}=1-\tau$ and 
\be
p_1^2\=m_1^2\=\tau^2M_0^2 + {\cal O}(k^2_\perp)\,, \qquad 
             p_2^2\=m_2^2\=\bar{\tau}^2M_0^2 + {\cal O}(k^2_\perp)\,.
\ee
The boost to the IMF leads to: 
\be
\bar{S}_0 \= \frac1{\sqrt{2}}\Big[\frac{2\xi}{1-\xi^2} \frac{k^2_\perp}{M_0}p\,\Sla
          - 2\frac{k^2_\perp}{1-\xi^2} + i\sigma^{\mu\nu} p_\mu K_\nu + M_0 K\Sla \Big]\,.
\label{eq:spin-wf}
\ee
For convenience the variable $\xi=1-2\tau$ is introduced. For $\xi=0$ this covariant spin 
\wf{} coincides with the one employed for the $\chi_{c0}$ in \ci{bolz97}. The normalization
of the spin \wf{} is chosen such that 
\be
{\rm Tr}\Big(S_0^\dagger S_0\Big)\= 4E^2k^2_\perp + {\cal O}(k^4_\perp)
\ee
where $E$ is the meson's energy. The meson's $s\bar{s}$ Fock state \req{eq:Fock} explicitly reads
\be
\langle f_0;p| \= \frac{\delta_{c\bar{c}}}{2\sqrt{N_c}}\int \frac{d\xi d^2k_\perp}{16\pi^3} 
                 \Psi_0(\xi,k^2_\perp)\bar{S}_0 
                      \langle s_c;p_1,\lambda_1| \langle \bar{s}_{\bar{c}};p_2,\lambda_1|\,,
\ee
The number of colors is denoted by $N_c$ and $c, \bar{c}$ are color labels.
Proper state normalization requires the condition
\be
 \frac12\int \frac{d\xi d^2k_\perp}{16\pi^3} k^2_\perp |\Psi_0(\tau,k^2_\perp)|^2=P_{f_0}\leq 1
\label{eq:probability}
\ee
where $P_{f_0}$ is the probability of the $s\bar{s}$ Fock component.

\subsection{Collinear reduction}
\label{sec:coll-reduction}
In collinear approximation the limit $k_\perp\to 0$ in the hard subprocess is to be taken
in general. However, terms $\propto K^\alpha$ in it combine with terms linear in $K$
in the spin \wf{} and therefore survive the $k_\perp$-integration of the \wf{}. These terms
are in general of the same order as the other terms in the spin \wf{} and it is
therefore unjustified to neglect these terms~\footnote{
  For $l=0$ hadrons these terms are suppressed by $k_\perp^2$; the leading term is $k_\perp$-independent.}.
Consider the expansion of the subprocess amplitude with respect to $K$:
\be
{\cal M} \= A_0(\xi) + K^\alpha A_{1\alpha}(\xi) + {\cal O}(K^\alpha K^\beta) 
\ee
where $A_0$ is of order 1 while $A_1$ is of order $1/p^+$ for dimensional reason. The
$k_\perp$-integration yields
\be
\int \frac{d^2k_\perp}{16\pi^3} \Psi_0(\xi,k^2_\perp) K\Sla\,{\cal M}
      \= -\frac12 g_\perp^{\nu\alpha} \gamma_\nu A_{1\alpha} \int 
                            \frac{dk^2_\perp}{16\pi^2}k^2_\perp \Psi_0\,.
\ee
Formally this is equivalent to the replacement 
\be
K\Sla \Longrightarrow -\frac{k^2_\perp}{2} g_\perp^{\nu\alpha} \gamma_\nu 
                            \frac{\partial}{\partial K_{\nu}}\Big|_{K\to 0}\,.
\ee         
The transverse metric tensor is defined by $g_\perp^{11}=g_\perp^{22}=-1$ while all other 
components are zero in a frame where the meson moves along the 3-direction. In the collinear 
limit the spin wave function becomes
\be
\bar{S}_0^{\rm coll} \= \frac{k^2_\perp}{\sqrt{2}}\,
                \Big[\frac{2\xi}{1-\xi^2}\frac{p\,\Sla}{M_0}
                   - \frac2{1-\xi^2}   
           - \frac12\big(i\sigma_{\mu\alpha} p^\mu + M_0\gamma_\alpha\big) g_\perp^{\alpha\beta}
                    \frac{\partial}{\partial K_{\beta}} \, \Big]_{K\to 0}\,.
\label{eq:spwf-coll-1}
\ee
Multiplying the spin \wf{} with the reduced \wf{} and integrating over $k_\perp$, 
one arrives at the associated \da s. The first term in \req{eq:spwf-coll-1} generates 
the twist-2 \da{}
\be
\frac{\bar{f}_0 }{2\sqrt{2N_c}}\Phi_0(\xi)\= \frac{2\xi}{1-\xi^2} 
           \int\frac{dk_\perp^2}{16\pi^2}\,\frac{k^2_\perp}{M_0}\,\Psi_0\,.
\label{eq:twist-2-da}
\ee
Because of charge conjugation invariance the twist-2 \da{} is antisymmetric in $\xi$. It 
possesses a Gegenbauer expansion and depends on the factorization scale, $\mu_F$, 
\ci{cheng05,CZ,lue06} 
\be
\Phi_0(\xi,\mu_F)\=\frac{N_c}{2} (1-\xi^2) \sum_{m=1,3,\dots}B_m(\mu_F)C_m^{3/2}(\xi)
\label{eq:twist-2-da-gegenbauer}
\ee
Evidently, the reduced \wf{} must be symmetric in $\xi$.
The Gegenbauer coefficients in \req{eq:twist-2-da-gegenbauer} which encode the soft, 
non-perturbative QCD, evolve with the factorization scale as
\be
B_m(\mu_F)\=B_m(\mu_0) \left(\frac{\als(\mu_0)}{\als(\mu_F)}\right)^{-\gamma_m/\beta_0}
\label{eq:Bm-evolve}
\ee
where
\be
\gamma_m\=C_F\Big(1 - \frac{2}{(m+1)(m+2)}+4\sum_{j=2}^{m+1}\frac1{j}\Big)\,.
\ee
Here, $\beta_0=(11N_c-2n_f)/3$, $C_F=4/3$ and $n_f$ denotes the number of active flavors.
For the initial scale, $\mu_0$, the value $1.41\,\gev$ is chosen in this article.
The decay constant $\bar{f}_0$ depends on the scale too \ci{cheng05}
\be
\bar{f}_0(\mu_F)\=\bar{f}_0(\mu_0) 
         \left(\frac{\als(\mu_0)}{\als(\mu_F)}\right)^{4/\beta_0}\,.
\label{eq:f0-evolve}
\ee

The other terms in \req{eq:spwf-coll-1} are of twist-3 nature although they do not 
correspond to the full twist-3 contributions since, in general, they also receive contributions
from a second reduced \wf{}. This is however of no relevance for the purpose of the present paper, 
namely the calculation of the $\gamma^* -f_0$ transition form factors. As we shall see in the 
following there is no twist-3 contribution to it. Anyway the $k_\perp$-integration of the other 
terms leads to two further distribution amplitudes which are related to $\Phi_0$ in the case at 
hand:
\be
\Phi_{0s}(\xi,\mu_F) \= \frac1{\xi} \Phi_0(\xi,\mu_F)\,, \qquad 
                \Phi_{0\sigma}(\xi,\mu_F)\=\frac{1-\xi^2}{4\xi} \Phi_0(\xi,\mu_F)\,.
\label{eq:twist-3-das}
\ee
Both these \da s are symmetric in $\xi$ and only the even terms appear in their Gegenbauer
expansions.

With the help of these \da s one can transform the product of \wf{} and collinear spin \wf{} 
\req{eq:spwf-coll-1}, integrated over $k_\perp$, into the form 
\ba
\int\frac{d^2k_\perp}{16\pi^3} \Psi_0(\xi,k^2_\perp)\bar{S}_0^{\rm coll} &=& 
     \frac{\bar{f}_0}{2\sqrt{2N_c}}\frac1{\sqrt{2}}\Big[
                    \Phi_0 p\,\Sla\, -  \Phi_{0s}M_0    \nn\\
          &-& \Phi_{0\sigma} M_0 (i\sigma_{\mu\alpha}p^\mu  
                                + M_0\gamma_\alpha)\;\, g_\perp^{\alpha\beta}
                 \frac{\partial}{\partial K_{\beta}} \Big]_{K \to 0}\,.
\label{eq:spwf-collinear}
\ea
This expression resembles the corresponding pion spin \wf{} to twist-3 accuracy 
\ci{beneke-feldmann,signatures}.

\subsection{A \wf{} for the $f_0$ meson}
\label{sec:wf}
For the evaluation of the transition form factors the light-cone \wf{} is to be specified.
It is modeled as a Gaussian in $k^2_\perp/(1-\xi^2)$ times the most general 
$\xi$ dependence
\be
\Psi_{0}\=c\sum_{n=0,2\ldots} \tilde{B}_n C_n^{3/2}(\xi)\exp{[-4\frac{a_0^2 k_\perp^2}{1-\xi^2}]} 
\label{eq:wf-general}
\ee
with
\be
c\=16\pi^2 \sqrt{2N_c} \bar{f}_0 M_0 a_0^4\,.
\label{eq:norm}
\ee
This \wf{} is similar to the one for the pion advocated for in \ci{BHL}. It has been used  
for instance in the calculation of the photon-pseudoscalar transition form factors 
\ci{kroll11} or in the analysis of pion electroproduction \ci{GK5}.
Insertion of the \wf{} into Eq.\ \req{eq:twist-2-da} leads to the associated \da{} 
\req{eq:twist-2-da-gegenbauer} with the Gegenbauer coefficients ($m$ is an odd integer)
\be
B_m\=\frac{m}{2m+1} \tilde{B}_{m-1} + \frac{m+3}{2m+5} \tilde{B}_{m+1}\,.
\ee

As a consequence of charge conjugation invariance which forces $\Psi_0$ to be
symmetric in $\xi$, the matrix element
\be
\langle f_0;p|\bar{s}(0)\gamma_\mu s(0)|0\rangle \= \frac{\sqrt{N_c}}{2} 
            \int d\xi \frac{dk^2_\perp}{16\pi^2} \Psi_0 {\rm Tr}\big[\bar{S}_0\gamma_\mu\big]
\ee
vanishes in accord with the result quoted in \ci{cheng05}. On the other hand, the scalar 
density provides
\be
\langle f_0;p|\bar{s}(0)s(0)|0\rangle \= M_0\bar{f}_0  
                   \= \frac{\sqrt{N_c}}{2} 
                      \int d\xi \frac{dk^2_\perp}{16\pi^2} \Psi_0 {\rm Tr}\big[\bar{S}_0\big]\,.
\label{eq:scalar-density}
\ee
Evaluation of the integral leads to $\tilde{B}_0\simeq -1$. This estimate is to be taken with 
caution since $\Phi_{0s}$ in \req{eq:twist-3-das} is likely not the full twist-3 \da{}, but it 
provides orientation. As is obvious from the vacuum-particle matrix element of quark field 
operators given in \req{eq:scalar-density}, the decay constant is a short-distance quantity; it 
represents the \wf{} at the origin of the configuration space. It is also clear that only the 
$s\bar{s}$ Fock component of the $f_0$-meson contributes to this matrix element.

For the numerical evaluation of the $\gamma^*-f_0$ transition form factors the \wf{}
will be restricted to the first Gegenbauer term, all others are neglected.
\be
\Psi_{01}\= 3cB_1\exp{[-4\frac{a_0^2 k_\perp^2}{1-\xi^2}]}
\label{eq:wf-example}
\ee
with 
\be
B_1\simeq \tilde{B}_0/3 \simeq -1/3\,.
\label{eq:gegenbauer-coefficients}
\ee
In this case the twist-2 \da{} reads  
\be
\Phi_{01}\=\frac{N_c}{2} (1-\xi^2) B_1 C_1^{3/2}(\xi)\,.
\label{eq:da-example}
\ee

For the transverse-size parameter, $a_0$, the value $0.8\,\gev^{-1}$ is taken
in the following. This value is very close to the corresponding value for the pion, see 
\ci{kroll11}. The r.m.s. $k_\perp$ is related to the transverse-size parameter by
\be
\sqrt{\langle k^2_\perp \rangle}\=\sqrt{\frac3{14}}\,\frac1{a_0}\,.
\ee
For the value $a_0=0.8\,\gev^{-1}$ the r.m.s. value of $k_\perp$ is $0.58\,\gev$ which is similar
to the corresponding results for the valence Fock components of other hadrons. 
For the decay constant the value
\be
\bar{f}_0(\mu_0)=(180\pm 15)\,\mev
\label{eq:fazio}
\ee
is adopted which has been derived by De Fazio and Pennington \ci{fazio} from radiative 
$\phi\to f_0\gamma$ decays with the help of QCD sum rules (see also \ci{bediaga}). In 
\ci{fazio} the $f_0$-meson is considered as a (dominantly) $s\bar{s}$ state. The value 
\req{eq:fazio} is extracted from the stability window for the Borel parameter between 
1.2 and $2\,\gev^2$. This is consistent with the initial scale chosen in this article.

\section{The $\gamma^* - f_0$ transition form factors}
\label{sec:form-factor}
\subsection{The definition of the form factors}
\label{sec:definition}
Let us consider the general case of two virtual photons
\be
\gamma^*(q_1,\lambda_1) + \gamma^*(q_2,\lambda_2)\to f_0(p)
\ee
where $q_i$ and $p$ denote the momenta of the photons and the mesons
while $\lambda_i$ are the helicities of the photons. One has
\be
q_1^2\=-Q_1^2\,, \qquad q_2^2\=-Q_2^2\,,  \qquad p^2=M_0^2\,.
\ee
It is convenient to introduce the following variables \ci{DKV1}
\be
\ov{Q}^2\=\frac12(Q_1^2+Q^2_2)\,, \qquad 
                   \omega\=\frac{Q^2_1 -  Q^2_2}{Q^2_1 + Q^2_2}
\label{eq:def}
\ee
where, obviously, $-1\leq \omega \leq 1$.

The transition vertex is defined by the matrix element of the time-ordered
product of two electromagnetic currents
\be
\Gamma^{\mu\nu}\=-ie_0^2 \int d^4x e^{-iq_1x}\langle f_0;p\mid 
             T\{j^\mu_{\rm em}(x)j^\nu_{\rm em}(0)\}\mid 0\rangle
\label{eq:vertex-def}
\ee
where
\be
j^\mu_{\rm em}\= e_u \bar{u}(x)\gamma^\mu u(x) + e_d \bar{d}(x)\gamma^\mu d(x) + 
                 e_s \bar{s}(x)\gamma^\mu s(x) 
\ee
and $e_i$ are the quark charges in units of the positron charge, $e_0$.
Following \ci{pascalutsa12} the vertex is covariantly decomposed as
\ba
\Gamma^{\mu\nu}&=&ie_0^2 \frac{q_1\cdot q_2}{M_0} \left\{\Big[-g^{\mu\nu}  
      + \frac1{\ov{Q}^4\kappa}
       \big[q_1\cdot q_2(q_1^\mu q_2^\nu+q_2^\mu q_1^\nu) \right. \nn\\ 
      &&\left.\hspace*{0.15\tw} +\, Q_1^2 q_2^\mu q_2^\nu + Q_2^2 q_1^\mu q_1^\nu\big)\big]\Big] 
       \,F_T(\ov{Q}^2,\omega)  \right.   \nn\\
    &-&\left. \frac{q_1\cdot q_2}{\ov{Q}^4\kappa}
      \Big[q_1^\mu + \frac{Q_1^2}{q_1\cdot q_2}q_2^\mu\Big]
    \Big[q_2^\nu + \frac{Q_2^2}{q_1\cdot q_2}q_1^\nu\Big]\,F_L(\ov{Q}^2,\omega) \right\}
\label{eq:vertex}
\ea
where 
\be
\kappa\=\frac{(q_1\cdot q_2)^2}{\ov{Q}^4} - 1 + \omega^2\= 
                 \omega^2 + \frac{M_0^2}{\ov{Q}^2}+\frac{M_0^4}{4\ov{Q}^4}\,.
\ee
Current conversation is manifest:
\be
q_{1\mu}\Gamma^{\mu\nu}\=0\,, \qquad   q_{2\nu}\Gamma^{\mu\nu}\=0\,.
\ee
As one sees from \req{eq:vertex} there are two form factors, one for transverse photon 
polarization, $F_T$, and another one for longitudinal polarization, $F_L$. By 
definition the form factors are dimensionless.

\begin{figure}[t]
\begin{center}
\includegraphics[width=0.5\tw]{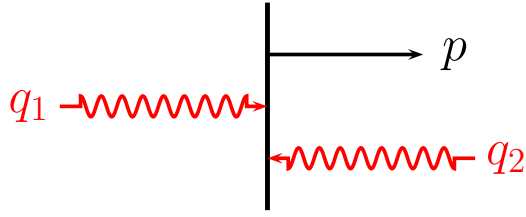}
\caption{\label{fig:1} The equal-energy brick wall frame.}
\end{center}
\end{figure}

Contracting the vertex function with the polarization vectors of the photons 
and using transversality  ($\eps_iq_i=0$), one arrives at 
\ba
\eps_1^\mu \eps_2^\nu\Gamma_{\mu\nu}&=& ie_0^2 \frac{q_1\cdot q_2}{M_0} 
               \left\{\Big[-\eps_1\cdot\eps_2 
          +  \frac{q_1\cdot q_2}{\kappa\ov{Q}^4}\eps_1\cdot q_2 \eps_2\cdot q_1\Big]\,F_T
              \right. \nn\\
         &&\left.\hspace*{0.1\tw} -\frac{1-\omega^2}{\kappa q_1\cdot q_2}
                   \eps_1\cdot q_2 \eps_2\cdot q_1\,F_L\right\}\,.
\label{eq:contractionFF1}
\ea 
One can show, most easily in the equal-energy brick wall frame (see Fig.\ \ref{fig:1}), 
defined by
\be
q_1\= (\nu\,0\,0\, a_1)\,, \qquad q_2\= (\nu\,0\,0\, a_2)\,, \qquad
                    p\=(2\nu\,0\,0\, a_1+a_2)\,,
\ee
that the contraction with transverse photon polarization vectors with the same
helicity projects out the form factor $F_T$ and with longitudinal ones $F_L$:
\ba
\eps_1^\mu(\lambda_1) \eps_2^\nu(\lambda_2)\Gamma_{\mu\nu} &=&
              -ie_0^2\frac{q_1\cdot q_2}{M_0} F_T(\ov{Q}^2,\omega)
                                         \delta_{\lambda_1\lambda_2}\,, \nn\\
\eps_1^\mu(0) \eps_2^\nu(0)\Gamma_{\mu\nu} &=&
      \phantom{-}ie_0^2\sqrt{1-\omega^2} 
                              \frac{\ov{Q}^2}{M_0}F_L(\ov{Q}^2,\omega)\,.
\label{eq:FF-BF}
\ea
If the photons have different helicities the vertex function is zero.

\subsection{The LO perturbative calculation}
\label{sec:LO}
\begin{figure}[t]
\begin{center}
\includegraphics[width=0.7\tw,bb=120 551 582 659]{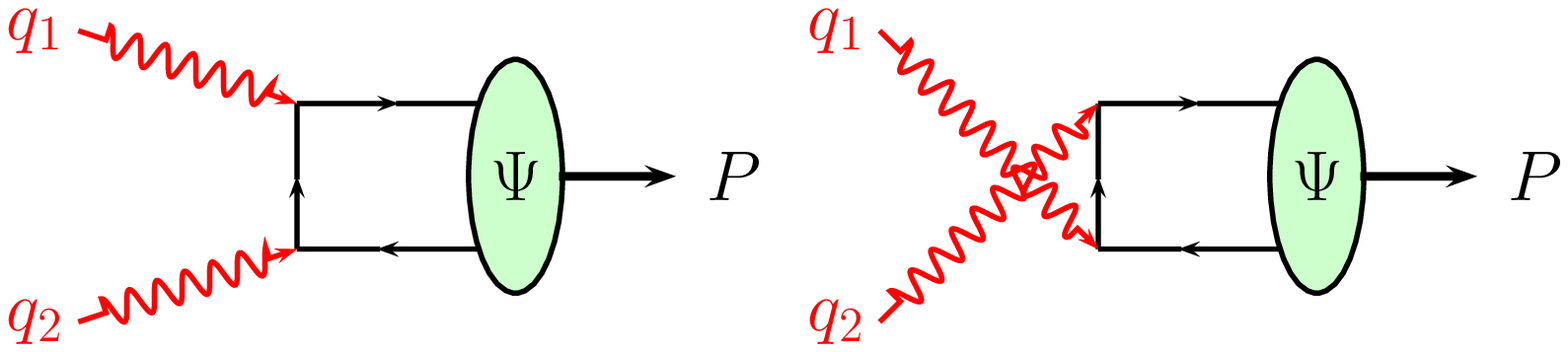}
\caption{\label{fig:2} LO Feynman graphs for the $\gamma^*\to f_0$
transition form factors. The momenta of the virtual partons are denoted by $g_1$ and $g_2$.}
\end{center}
\end{figure}

In the perturbative calculation of the form factors, performed at large $\ov{Q}^2$, 
the mass of the $f_0$-meson is neglected whenever this is possible. From the Feynman 
graphs shown in Fig.\ \ref{fig:2} one finds for the vertex function \req{eq:vertex-def} 
\ba
\Gamma_{\mu\nu}&=&-i \frac12 e_0^2e_s^2 \sqrt{N_c} 
              \int \frac{d\xi d^2k_\perp}{16\pi^3}\Psi_0(\xi,k_\perp)\left\{ 
               {\rm Tr}\Big[\bar{S}_0 \,\gamma_\mu 
         \frac{\frac12(1-\xi) p\,\Sla + K\Sla -q_1\,\sla}{g_1^2}\,\gamma_\nu\Big]\right. \nn\\
&& \left. \hspace*{0.25\tw}+{\rm Tr}\Big[\bar{S}_0 \,\gamma_\nu 
         \frac{\frac12(1-\xi) p\,\Sla + K\Sla - q_2\,\sla}{g_2^2}\, \gamma_\mu\Big] \right\}
\ea
where the parton virtualities read (see also Fig.\ \ref{fig:2})
\be
g_1^2\= -\ov{Q}^2(1+\xi\omega)-k_\perp^2\,, \qquad g_2^2\= -\ov{Q}^2(1-\xi\omega)-k_\perp^2\,.
\ee
Taking into consideration that the traces are only non-zero for even numbers of $\gamma$ 
matrices, one notices that only the first term of the spin \wf{} \req{eq:spin-wf}, i.e.\ the 
leading-twist piece, contributes to the traces. The twist-3 terms lead to an odd number
of $\gamma$ matrices in the traces, the fourth term is neglected. With the help of \req{eq:FF-BF} 
one finally arrives at the following expressions for the form factors:
\ba
F_T(\ov{Q}^2,\omega)&=&  - 4\sqrt{2N_c}\frac{e_s^2}{\ov{Q}^2} 
       \frac{\omega^2 + \frac12\frac{M_0^2}{\ov{Q}^2}}{1+\frac12\frac{M_0^2}{\ov{Q}^2}} 
           \int \frac{d\xi dk^2_\perp}{16\pi^2}k^2_\perp \Psi_0(\xi,k_\perp)
                  \frac{\xi^2}{1-\xi^2}\,  \nn\\
        &&   \hspace*{0.25\tw} \times \frac1{1-\xi^2\omega^2 + 2 k^2_\perp/\ov{Q}^2}\,, \nn\\ 
F_L(\ov{Q}^2,\omega)&=& -\frac12 \frac{M_0^2}{\ov{Q}^2} 
           \frac{1 + \frac12\frac{M_0^2}{\ov{Q}^2}}{\omega^2+\frac12\frac{M_0^2}{\ov{Q}^2}}
             F_T(\ov{Q}^2,\omega)\,. 
\label{eq:FF}
\ea

Because of the variation of $\omega^2$ between 0 and 1 the mass dependent terms in front
of the integral are kept. For $\omega\to \pm 1$ they exactly cancel whereas
for $\omega\to 0$ $F_T\propto \ov{Q}^{-4}$. The \wf{} \req{eq:wf-general} generates a factor 
$(1-\xi^2)^2$ in the $k^2_\perp$ integration. Hence, there is no singularity at the end points 
$\xi\to \pm 1$ for all $\omega$. One also notices from \req{eq:FF} that
\be
F_{T,L}(\ov{Q}^2,-\omega)\= F_{T,L}(\ov{Q}^2,\omega)\,.   
\ee
For $\omega\gg M_0^2/(2\ov{Q}^2)$ the terms $\sim M_0^2/\ov{Q}^2$ in \req{eq:FF} can be
neglected  and 
\be
F_T \propto 1/\ov{Q}^2\,,  \qquad  F_L \propto 1/\ov{Q}^4\,.
\ee
For $\omega \to 0$, on the other hand,  only the term $\sim M_0^2/(2\ov{Q}^2)$  remains and
\be
F_{T,L} \propto 1/\ov{Q}^4\,.
\ee

Explicitly, for $\omega\to 1$  (i.e.\ $Q_2^2=0$)
\ba
F_T(Q_1^2,1)&=&-8\sqrt{2N_c}\frac{e_s^2}{Q_1^2} \int \frac{d\xi dk^2_\perp}{16\pi^2}
                          k^2_\perp \Psi_0(\xi,k_\perp)\nn\\
          &\times& \frac{\xi^2}{1-\xi^2}\, \frac1{1-\xi^2+4k^2_\perp/Q_1^2}\,. 
\label{eq:mod-pert-FF}
\ea

\begin{figure}[t]
\begin{center}
\includegraphics[width=0.5\tw,bb=106 303 558 655]{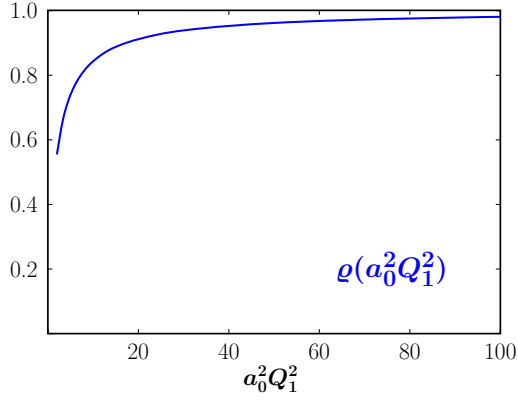}
\caption{\label{fig:reduction} The reduction function $\varrho$ versus $a_0^2Q_1^2$.}
\end{center}
\end{figure} 
For a wave function of the type \req{eq:wf-general} one can write Eq.\ \req{eq:mod-pert-FF} 
as
\be
F_T(Q_1^2,1)\= \varrho(a_0^2Q_1^2)\,  F_T^{\rm coll}(Q_1^2,1) 
\label{eq:FT-fact}
\ee
with
\ba
F_T^{\rm coll}&=&-2\frac{e_s^2}{Q_1^2} \bar{f}_0M_0\int d\xi\frac{\xi\Phi_0(\xi)}{1-\xi^2} \nn\\
              &=& -2N_c\frac{e_s^2}{Q_1^2} \bar{f}_0M_0\sum_{m=1,3\ldots} B_m 
\label{eq:FT-coll}
\ea
and 
\be
\varrho(x)\=\int dK\frac{K e^{-K}}{1+K/x}\,.
\ee
For this type of \wf s the transition form factor is given by the collinear result multiplied
by a universal reduction factor $\varrho$. The latter function is shown in Fig.\ 
\ref{fig:reduction}. It is interesting that, in the collinear approximation, the LO 
perturbative result for the form factor is related to the sum over all Gegenbauer coefficients.
The $\gamma-\pi$ transition form factor possesses this property too. This makes it clear
that it is impossible to extract more than one Gegenbauer coefficient from the $\gamma-f_0$ 
transition form factor data. This coefficient is to be regarded as an effective one. NLO 
corrections may allow one to fix a second coefficient \ci{DKV1}. The situation improves for 
$|\omega| < 1$ as will be discussed in Sect.\ \ref{sec:two-virtualities}. 

\section{Results}
\label{sec:results}
\subsection{The real-photon limit}
\label{sec:real-photon}
The BELLE collaboration \ci{belle15} extracted the $\gamma - f_0$ transition form factor 
from the cross sections on $\gamma^*\gamma\to \pi^0\pi^0$. In order to fix the normalization 
of that form factor the couplings of the $f_0$ to both the two-photon and the $\pi\pi$ 
channels are required. Both these couplings are not well known \ci{PDG}. Hence, the 
normalization of the $\gamma - f_0$ transition form factor is subject to considerable 
uncertainties. The published data on the transition form factor, $F_T(Q_1^2)$, are scaled by 
the value of the form factor at $Q_1^2=0$ obtained from the width of the two-photon decay 
of the $f_0$-meson ($M_0=(990\pm 20)\,\mev$ \ci{PDG})
\be
\Gamma(f_0\to\gamma\gamma)\=\frac{\pi}{4}\ale^2 M_0 |F_T(0)|^2\,.
\ee 
From the average decay width quoted in \ci{PDG}, one obtains
\be
|F_T(0)|\= 0.0865 \pm 0.0141\,.
\label{eq:FT0}
\ee
The BELLE collaboration uses the slightly different value 
$|F_T(0)|_{\rm BELLE}\=0.0832 \pm 0.0136$.

\begin{figure}[t]
\begin{center}
\includegraphics[width=0.6\tw,bb=109 378 586 718]{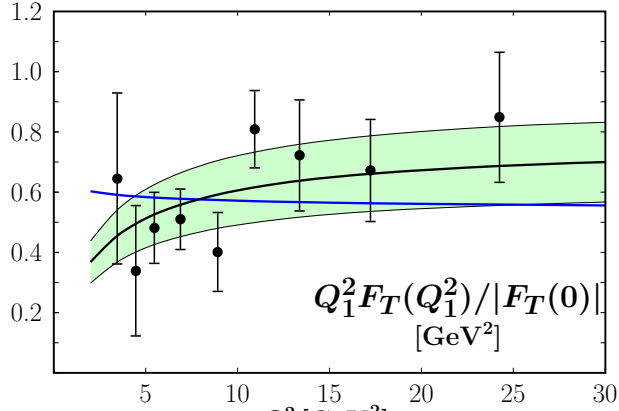}
\caption{\label{fig:transition-FF} The $Q^2$-dependence of the $\gamma - f_0$ transition 
form factor scaled by $|F_T(0)|/Q_1^2$ (for $|F_T(0)|$ the value \req{eq:FT0} is taken).
Data are taken from \ci{belle15}; only the statistical errors are shown. 
The dashed and solid  lines are the results of the collinear approximation and the modified 
perturbative approach evaluated from \wf{} \req{eq:wf-example}, respectively. The shaded band 
represents the normalization uncertainty of the second result.} 
\end{center}
\end{figure}

In a first step the BELLE data are compared to the collinear result \req{eq:FT-coll} for $F_T$. 
For the factorization scale $\muF=Q_1^2$ is used and for $\LQCD$ the value $180\,\mev$ in 
combination with four flavors. Allowing only for the first Gegenbauer term in the expansion 
\req{eq:twist-2-da-gegenbauer} of $\Phi_0$ and taking for the decay constant the value 
\req{eq:fazio}, we fit $B_1$ against the BELLE data. The fit yields $B_1^{\rm coll}(\mu_0)=-0.44\pm 0.04$ 
and $\chi^2=10.3$ for 9 data points. The fitted value of $B_1$ is not far from the estimate
quoted in \req{eq:gegenbauer-coefficients}. For these \wf{} parameters the probability of the 
$s\bar{s}$ Fock component of the $f_0$-meson is (see\req{eq:probability}):
\be
P_{f_0}\=\frac{12}{5}N_c\big[\pi\bar{f}_0M_0a_0^2 B_1\big]^2\=0.18\,.
\ee
The results of the fit are shown in Fig.\ \ref{fig:transition-FF}. Reasonable agreement with 
experiment is to be seen within rather large errors although the shape of the fit is opposite 
to that of the data: the collinear result for the scaled form factor, $Q_1^2F_T^{\rm coll}$ 
slightly decreases with increasing $Q_1$ due to the evolution of the decay constant and the 
Gegenbauer coefficient, $B_1$, whereas the data increase in tendency. 

An increasing scaled form factor can be generated by quark transverse momenta in the hard 
scattering kernel and in the \wf{}, see Fig.\ \ref{fig:reduction}. Retaining the quark 
transverse momenta implies that quarks and antiquarks are pulled apart in the transverse 
configuration or impact-parameter space. The separation of color sources is accompanied by 
the radiation of gluons. These radiative corrections have been calculated in Ref.\ \ci{li-sterman}
in the form of a Sudakov factor in the impact parameter plane. The Sudakov factor, $e^{-S}$, 
comprises resummed leading and next-to-leading logarithms which are not taken into account 
by the usual QCD evolution. The $k_\perp$-factorization combined with the Sudakov factor is 
termed the modified perturbative approach (mpa)\ci{li-sterman}. It has been used, for instance, 
in calculations of the pion electromagnetic form factor \ci{li-sterman} or the $\pi -\gamma$ 
transition form factor \ci{kroll11} and will be used here as well. In the impact-parameter 
plane the transition form factor \req{eq:mod-pert-FF} reads  
\ba
F_T(Q_1^2,1) &=&-\frac{e_s^2\sqrt{2N_C}}{2\pi}\,\int_{-1}^1 d\xi 
                      \frac{\xi^2}{1-\xi^2}\nn\\
        &\times& \int_0^{1/\LQCD} db b \Big[k_\perp^2\Psi_0\Big]\,e^{-S} 
                      K_0\Big(bQ_1/2\sqrt{1-\xi^2}\,\Big)\,.
\label{eq:FT-impact}
\ea
The integrand is completed by the Sudakov factor, $\exp{(-S)}$, its explicit form can
be found for instance in \ci{kroll11}. The Sudakov factor provides the sharp cut-off
of the $b$-integral at $1/\LQCD$. Since $1/b$ in the Sudakov factor marks the interface
between the non-perturbative soft momenta which are implicitly accounted for in the meson
\wf{}, and the contributions from soft gluons, incorporated in a perturbative way in
the Sudakov factor \ci{li-sterman,kroll11}, it naturally acts as the factorization scale.
The Bessel function $K_0$ is the Fourier transform of the hard scattering kernel and 
$\Big[k^2_\perp \Psi_0\Big]$ is the Fourier transform of the \wf{} \req{eq:wf-example} 
multiplied by $k^2_\perp$. It reads
\ba
\Big[k^2_\perp\Psi_0\Big]&=& \frac{3\pi}{4} \sqrt{2N_C}\bar{f}_0 M_0 B_1(1-\xi^2)^2
                  \Big(1-\frac{1-\xi^2}{16a_0^2}b^2\Big)e^{-\frac{1-\xi^2}{16a_0^2}b^2}\,.
\label{eq:wf-b}
\ea
Evaluating the form factor within the modified perturbative approach and fitting $B_1$
to the BELLE data \ci{belle15} one arrives at the results shown in Fig.\ \ref{fig:transition-FF}.
The fit provides the following value for the Gegenbauer coefficient~\footnote{
 As shown for the case of the $\gamma - \pi$ form factor in \ci{kroll11} the contributions
 from the higher Gegenbauer terms are suppressed as compared to the lowest one. This
 property of the modified perturbative approach comes into effect here, too.}
\be
B_1^{\rm mpa}(\mu_0)\=-0.57\pm 0.05
\label{eq:B1-mpa}
\ee
and $\chi^2=5.9$ for 9 data points. The normalization uncertainty of the theoretical
result follows from the errors of $B_1$ and $F_T(0)$, see \req{eq:FT0}. The agreement
of the result obtained within the modified perturbative approach, with experiment is somewhat 
better than for the collinear approximation - the scaled form factor increases with $Q^2_1$ 
as the data do. This increase is the effect of the $k_\perp$ corrections shown in 
Fig.\ \ref{fig:reduction}, the Sudakov factor plays a minor role in this context~\footnote{
  In the analysis of the $\gamma - f_2$ form factor performed in \ci{braun16} the collinear
  factorization framework does also not suffice. In order to achieve fair agreement with
  experiment \ci{belle15} soft end-point corrections have to be included in the analysis.}.
In passing it is noted that the predictions presented in \ci{schuler97} lie markedly
below experiment for $Q_1^2\gsim 10\,\gev^2$.

The value \req{eq:B1-mpa} of the Gegenbauer coefficient $B_1$ is not far from the 
QCD sum result \ci{cheng05}:
\be
\bar{f}_0(\mu_0)\=(410\pm 22)\,\mev\,, \qquad B_1(\mu_0)\=-0.65\pm 0.07\,.
\label{eq:cheng}
\ee
The coefficent $B_3$ is found to be zero within errors. However, the value of $\bar{f}_0$
is substantially larger than the value \req{eq:fazio} used in the form factor calculation.
More precisely, the fit to the BELLE data fixes the product of $\bar{f}_0$ and $B_1$
for which the following results exist 
\ba
\bar{f}_0(\mu_0)B_1(\mu_0) &=& (-0.079 \pm 0.007)\,\gev \hspace*{0.1\tw} {\rm collinear}\nn\\
                             &=& (-0.103 \pm 0.990)\,\gev \hspace*{0.1\tw} 
                                                             {\rm mpa}\nn\\ 
                             &=& (-0.267 \pm 0.029)\,\gev \hspace*{0.1\tw} \ci{cheng05}
\ea
The product of $\bar{f}_0$ and $B_1$ derived in \ci{cheng05} is substantially larger 
than the BELLE data \ci{belle15} on the $\gamma - f_0$ transition form factors allow.
This product of $\bar{f}_0$ and $B_1$ is also in conflict with a light-cone \wf{} interpretation
since it leads to a probability larger than 1. Of course a smaller value of the 
transverse-size parameter would cure this problem for the prize of an implausible compact 
valence Fock component. For instance, if one halves $a_0$ the probability is about 0.12 but 
$\sqrt{\langle k^2_\perp\rangle}\simeq 1.2\,\gev$.

The last issue to be discussed is the contribution from the non-strange $q\bar{q}$ Fock state  
to the $\gamma - f_0$ transition form factor. This is usually considered 
as $f_0 - \sigma$ mixing \ci{ochs}-\ci{hooft}. As for the $\eta - \eta'$ system \ci{FKS1} this
mixing is treated in the quark-flavor basis. As a consequence of the smallness of OZI-rule
violations $\eta - \eta'$ mixing is particularly simple in that basis - there is a common mixing
angle for the states and the decay constants. It is assumed that this mixing scheme also holds 
for the case of interest here. Let $\sigma_n$ and $\sigma_s$ be states with the lowest Fock 
components $n\bar{n}=(u\bar{u}+d\bar{d})/\sqrt{2}$ and $s\bar{s}$, respectively. In analogy to
\req{eq:scalar-density} the corresponding decay constants are defined by the $\sigma_i$-vacuum 
matrix elements of the quark field operators:
\be
\langle \sigma_n|\bar{n}(0)n(0)|0\rangle=M_{\sigma_n}\bar{f}_n\,,  \qquad
\langle \sigma_s|\bar{s}(0)s(0)|0\rangle=M_{\sigma_s}\bar{f}_s\,.
\ee
Since in hard processes only small spatial quark-antiquark separations are of relevance
it seems plausible to embed the particle dependence and the mixing behavior of the $q\bar{q}$
Fock components solely into the decay constants~\footnote{
I.e.\ with the exception of the decay constants, the \wf s of the basis states, $\sigma_n$ and
$\sigma_s$, are assumed to be the same.}
(for a detailed discussion of this procedure in the $\eta - \eta'$ case see \ci{passek03}).  
In generalization of \req{eq:scalar-density} one may also define the decay constants 
$\bar{f}_i^q$ ($i=f_0,\sigma$; $q=n,s$) 
\be
\langle i | \bar{q}(0)q(0)|0\rangle \= M_i \bar{f}_i^q\,.
\ee
These decay constants mix according to 
\ba
\bar{f}_\sigma^n&=&\bar{f}_n\cos{\varphi}\,, \qquad \bar{f}_\sigma^s\=-\bar{f}_s\sin{\varphi}\,,\nn\\
\bar{f}_0^n&=&\bar{f}_n\sin{\varphi}\,, \qquad \bar{f}_0^s\=\phantom{-}\bar{f}_s\cos{\varphi}\,.
\ea
Hence, the $\gamma^* - f_0$ transition form factors are made of two contributions
\be
    F_{T,L} \= F_{T,L}^n + F_{T,L}^s
\label{eq:sum}
\ee
where the $n$ and $s$ contributions differ from \req{eq:FF} only by the decay constants, $\bar{f}_n$ 
and $\bar{f}_s$, the mixing angle, $\varphi$, and the quark charges, $(e_u^2+e_d^2)/\sqrt{2}$ and 
$e_s^2$. Thus, the contribution from the $n\bar{n}$ Fock state is taken into account if in 
\req{eq:FF}, and in other expressions derived for the form factors, the decay constant, $\bar{f}_0$,
is to be replaced by an effective one defined by
\be
\bar{f}^{\rm eff}_0\=\bar{f}_n\sin{\varphi}\frac1{\sqrt{2}}\frac{e_u^2+e_d^2}{e_s^2}
                       + \bar{f}_s\cos{\varphi}\,.
\label{eq:effective}
\ee
According to \ci{cheng05,cheng04} $\bar{f}_n\simeq \bar{f}_s$. Since the decay 
constant quoted in \req{eq:fazio} is to be identified with $\bar{f}_0^s$
and since $|\cos{\varphi}|$ is close to 1, see \req{eq:mixing}, it suffices
to assume  $\bar{f}_n\simeq \bar{f}_s\simeq \bar{f}_0$ for a rough estimate.
For the range $\varphi=(25 - 40)^\circ$ of the mixing angle quoted in \req{eq:mixing} 
one finds  
\be
\bar{f}^{\rm eff}_0/\bar{f}_0 \= 2.4 - 3.0\,.
\ee
Clearly, this leads to a transition form factor which is in conflict with the BELLE data
\ci{belle15}. Using the second range of mixing angles in \req{eq:mixing} one obtains 
reasonable agreement with experiment. Particularly favored is the range 
$\varphi=(145 - 151)^\circ$ for which the form factor stays within the uncertainty band 
displayed in Fig.\ \ref{fig:transition-FF}. An exact determination of the mixing angle is 
not possible at present given the poor information available for the basic decay constants, 
$\bar{f}_n, \bar{f}_s$, and the assumption on the explicit form of the light-cone \wf.

\subsection{The case of two virtual photons}
\label{sec:two-virtualities}
\begin{figure}[t]
\begin{center}
\includegraphics[width=0.6\tw,bb=102 395 494 674]{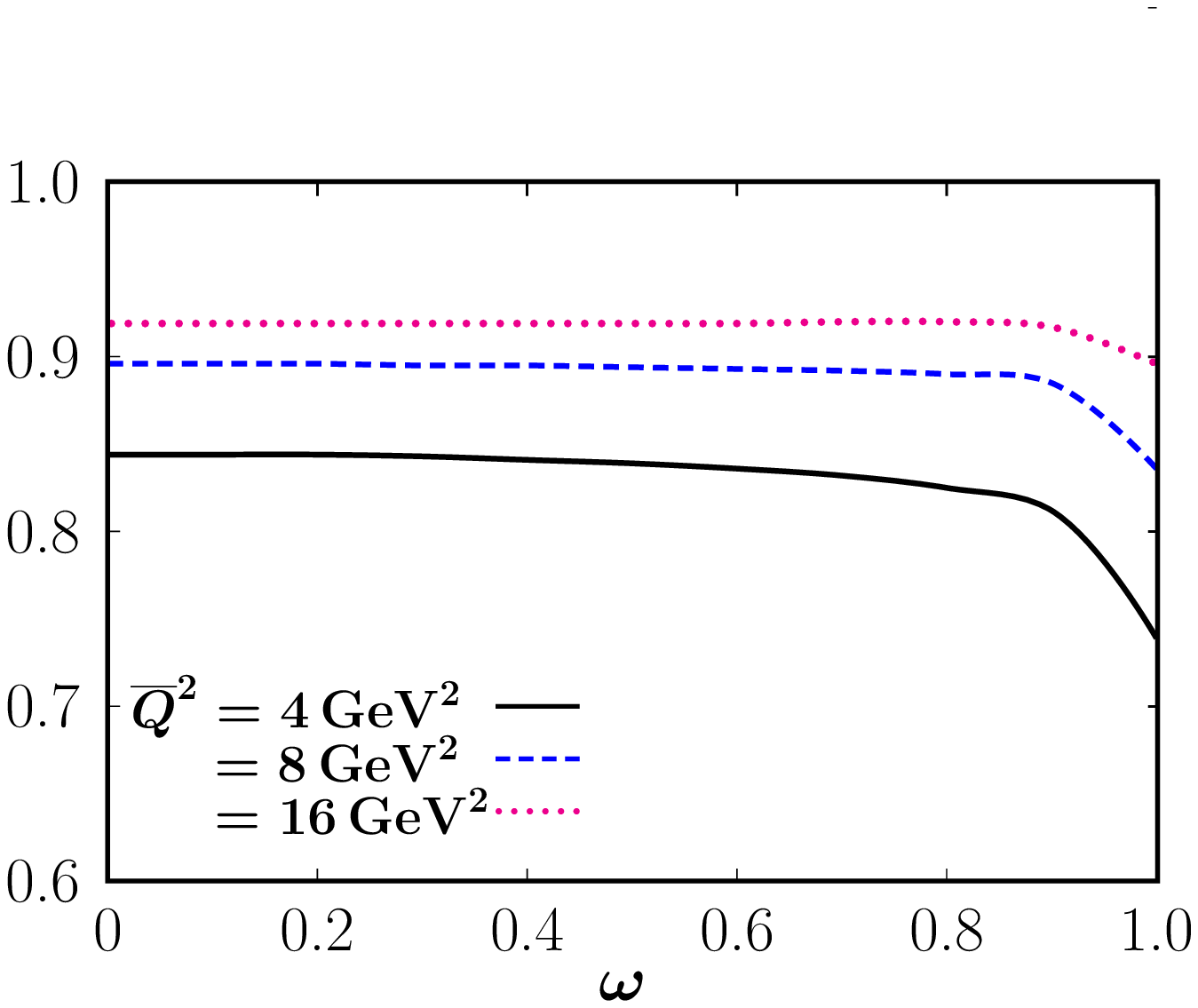}
\caption{\label{fig:ratio} The ratio of the transition form factors
evaluated from \req{eq:FF} and from the collinear result \req{eq:coll-omega}
versus $\omega$ for a set of $\ov{Q}^2$ values. The form factors are evaluated
from the \wf{} \req{eq:wf-example} and the associated \da{} \req{eq:da-example}, respectively.}
\end{center}
\end{figure} 

Here, in this subsection, it will be commented on the $\gamma^* - f_0$ transition
form factor. As is the case for $\omega=1$, the Sudakov factor plays a minor role.
In order to estimate the importance of the power corrections taken into account in the
modified perturbative approach the ratio of the form factors evaluated from \req{eq:FF}
(transformed to the impact parameter plane and with the Sudakov factor included) 
and from the collinear approximation 
\be
F_T^{\rm coll}(\ov{Q}^2,\omega)\= -\frac{e_s^2}{\ov{Q}^2} \bar{f}_0M_0 
               \frac{\omega^2 + \frac12\frac{M_0^2}{\ov{Q}^2}}{1+\frac12\frac{M_0^2}{\ov{Q}^2}}
           \int d\xi \frac{\xi \Phi_0(\xi)}{1-\xi^2\omega^2}
\label{eq:coll-omega}
\ee
is displayed in Fig.\ \ref{fig:ratio}. As expected the power corrections become smaller with
increasing $\ov{Q}^2$ and their importance decreases if $\omega$ deviates from 1.
The same observation has been made in \ci{DKV1} in case of the $\gamma^* - \pi$ transition
form factor. As noticed in \ci{DKV1} the reason for this effect is the term $1-\xi^2\omega^2$
in the hard scattering kernel which controls to which extent the form factor is sensitive
to contributions from the end-point regions $\xi \to \pm 1$ where soft effects can
be important.   

Since the power corrections are small at small $\omega$, it is of interest to look
at the transition form factor \req{eq:coll-omega} in this region. Using the Gegenbauer 
expansion of the \da{} the integral can be carried out term by term. The full result is a power 
series in $\omega^2$ leaving aside the $\omega$-dependence of the prefactor. The first terms of 
this series read
\ba
F_T^{\rm coll}(\ov{Q}^2,\omega)&=&-\frac25N_c e_s^2 \frac{\bar{f}_0 M_0}{\ov{Q}^2}
       \frac{\omega^2 + \frac12\frac{M_0^2}{\ov{Q}^2}}{1+\frac12\frac{M_0^2}{\ov{Q}^2}}
              \Big[B_1 + \omega^2\frac37\big(B_1+\frac{20}{27} B_3\big) \nn \\
   &+&   \omega^4\frac{5}{21} \big(B_1 + \frac{40}{33} B_3 + \frac{56}{143} B_5\big)
                    + \ldots \Big]\,.
\label{eq:coll-omega-expansion}
\ea 
As one notices the $m$-th Gegenbauer coefficient comes with the power $\omega^{m-1}$ first.
For $\ov{Q}^2$ larger than $4\,\gev^2$ the difference between the modified perturbative 
approach and the collinear result is smaller than $10\%$. Hence, the result in the modified 
perturbative approach evaluated from the \wf{} \req{eq:wf-general}, is not far from the collinear 
result \req{eq:coll-omega-expansion}. Thus, as is the case for the $\gamma^* - \pi$ transition 
form factor \ci{DKV1}, a measurement of the $\gamma^* - f_0$ transition form factors for a range 
of small $\omega$ would therefore provide valuable constraints on the $f_0$ \da{}.

\begin{figure}[t]
\begin{center}
\includegraphics[width=0.6\tw,bb=114 394 529 710]{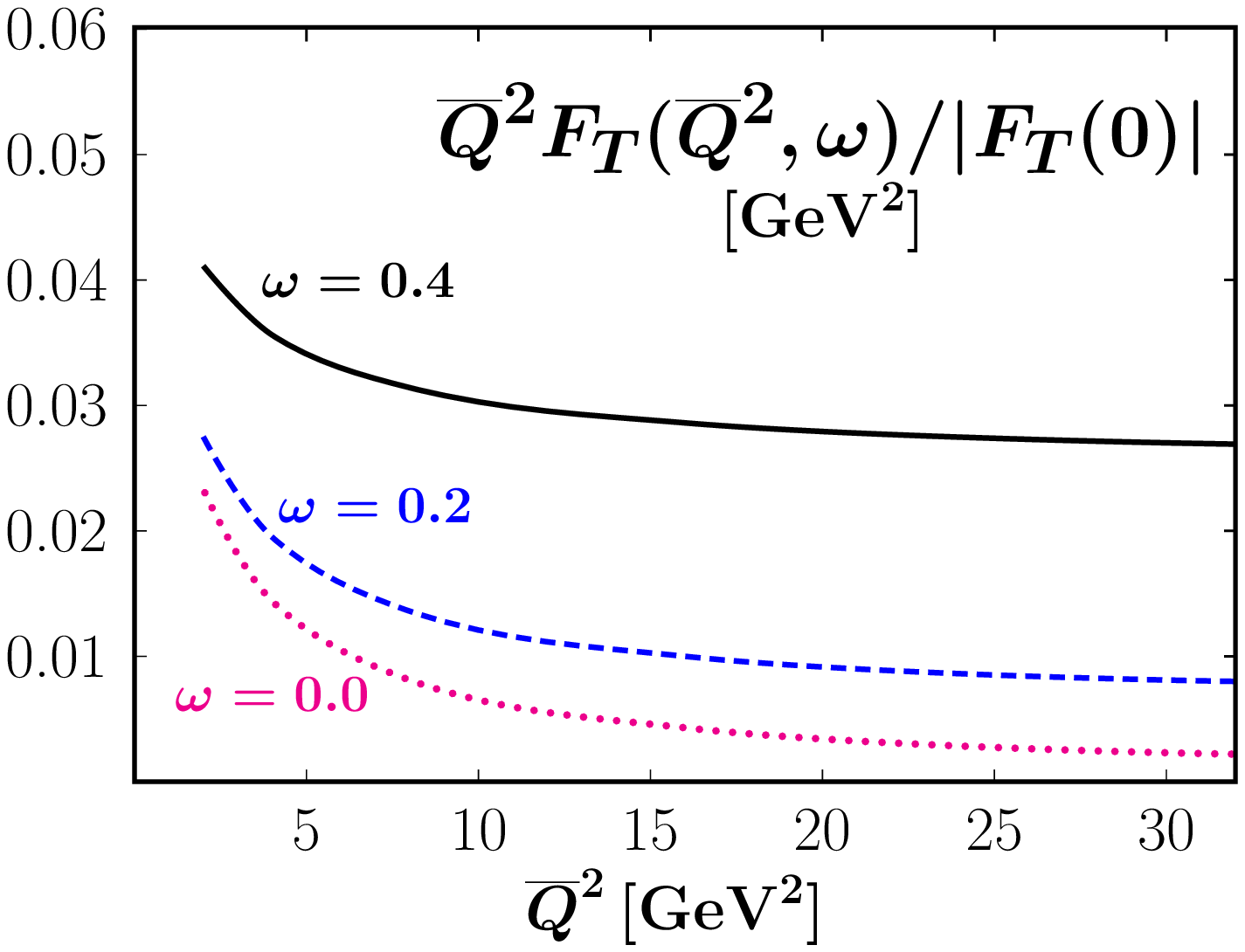}
\caption{\label{fig:omega} The $\gamma^* - f_0$ transition form factor, scaled by
$|F_T(0)|/\ov{Q}^2$, evaluated from the \wf{} \req{eq:wf-example} (with $B_1=-0.57$) within 
the modified perturbative approach versus $\ov{Q}^2$ for a set of $\omega$ values.}
\end{center}
\end{figure} 
In Fig.\ \ref{fig:omega} the $\gamma^*- f_0$ transition form factor, evaluated from
the \wf{} \req{eq:wf-example} within the modified perturbative approach, is shown for several
small values of $\omega$. It is clearly seen that the form factor drops with $\ov{Q}^2$
increasingly stronger than $1/\ov{Q}^2$ with decreasing $\omega$. At $\omega=0$
it decreases as $1/\ov{Q}^4$ (aside from evolution logarithms).

\section{Summary}
In this article the spin \wf{} of the $f_0(980)$ meson is constructed under the assumption 
that the meson is dominantly a strange-antistrange quark state. The collinear limit of the spin \wf{}
is also discussed and the connection to the twist-2 and twist-3 \da s is made. The spin \wf{} 
is applied in a calculation of the $\gamma^* - f_0$ transition form factors. In the real-photon 
limit the results for the transverse form factor are compared to the large-$Q^2$ 
data measured by the BELLE collaboration recently. It turns out that, for the $Q^2$
range explored by BELLE, the collinear approximation does not suffice, power corrections
to it, modeled as quark transverse moment effects, seem to be needed. The parameters required
in this calculation in order to achieve agreement with BELLE form factor data,
the transverse-size parameter, $a_0$, the decay constant, $\bar{f}_0$, and the lowest
(effective) Gegenbauer coefficient, $B_1$, have plausible values. However, Cheng {\it et al}
\ci{cheng05} in their analysis of charmless $B$-meson decays, adopt a much larger value
for $\bar{f}_0$ than \req{eq:fazio}. It remains to be seen whether the $B$-meson decays
can be reconciled with the decay constant \req{eq:fazio}.
The implications of $\sigma - f_0$ mixing for the transition form factors are also briefly 
discussed. A mixing angle of about $150^\circ$ seems to be favored.
The paper is completed by presenting results on the $\gamma^* - f_0$ form factors and
on their collinear limits. It turns out that, in many aspects, the photon - $f_0$ 
form factors have properties similar to the form factors for the transition from a photon to 
the $\pi^0$ or other pseudoscalar mesons. However, the limits for $Q_1^2\to\infty$ are 
different. Whereas for the pseudoscalar mesons the limits of the scaled form factors are
finite (e.g. $Q_1^2F_{\gamma \pi^0}\to \sqrt{2}f_\pi$) the $\gamma - f_0$ form factor $F_T$ 
tends to zero $\sim f_0(\mu_0)B_1(\mu_0)(\alpha_s(\mu_0)/\alpha_s(Q_1^2))^{-4/25}$. 
The $\gamma^* - f_0$ transition form factors also play a role in the calculation of the 
hadronic light-by-light contribution to the muon anomalous magnetic moment 
\ci{pauk14} - \ci{jegerlehner15}. In particular, the results presented in this article clarify
the asymptotic behavior of the $\gamma^* - f_0$ form factors.\\ 

{\it Acknowledgements:} Thanks to Volodya Braun and Andreas Sch\"afer for suggesting this study
and for discussions. The work is supported in part by the BMBF, contract number 05P12WRFTE.

\end{document}